\documentclass[conference]{IEEEtran}

\IEEEoverridecommandlockouts
\usepackage{cite}
\usepackage{amsmath,amssymb,amsfonts}
\usepackage{algorithmic}
\usepackage{framed}
\usepackage{graphicx}
\usepackage{tabularx}
\usepackage{multirow} 
\usepackage{wrapfig}
\usepackage{textcomp}
\usepackage{xcolor}
 \usepackage[colorlinks=true,
        linkcolor=black,
        citecolor=black,
        filecolor=black,
        urlcolor=black,
        bookmarks=true,
        bookmarksopen=true,
        bookmarksopenlevel=3,
        plainpages=false,
        pdfpagelabels=true]{hyperref}
\def\BibTeX{{\rm B\kern-.05em{\sc i\kern-.025em b}\kern-.08em
    T\kern-.1667em\lower.7ex\hbox{E}\kern-.125emX}}

\begin{document}

\title{When Less is More: A systematic review of
four-day workweek conceptualizations and their effects on organizational
performance} 

\author{
\IEEEauthorblockN{Marvin Auf der Landwehr}
\IEEEauthorblockA{University of Applied Sciences Hannover \\
\textit{Dpt. of Business Information Systems}\\
Hannover, Germany \\
marvin.auf-der-landwehr@hs-hannover.de}
\and
\IEEEauthorblockN{Julia Topp}
\IEEEauthorblockA{Sennheiser AG\\
\textit{Product Development}\\
Hannover, Germany \\
julia.topp@sennheiser.de}
\and
\IEEEauthorblockN{Michael Neumann}
\IEEEauthorblockA{University of Applied Sciences Hannover \\
\textit{Dpt. of Business Information Systems}\\
Hannover, Germany \\
michael.neumann@hs-hannover.de}
}

%\author{
%\IEEEauthorblockN{Anonymized Author}
%\IEEEauthorblockA{Anonymized University  \\
%\textit{Anonymized Department}\\
%City, Country \\
%firstname.lastname@email.org}
%\and
%\IEEEauthorblockN{Anonymized Author}
%\IEEEauthorblockA{Anonymized University  \\
%\textit{Anonymized Department}\\
%City, Country \\
%firstname.lastname@email.org}
%\and
%\IEEEauthorblockN{Anonymized Author}
%\IEEEauthorblockA{Anonymized University  \\
%\textit{Anonymized Department}\\
%City, Country \\
%firstname.lastname@email.org}
%}

%\author{\IEEEauthorblockN{Anonymized authors}
%\IEEEauthorblockA{Anonymized University \\
%\textit{Anonymized department}\\
%City, Country \\
%anonymized email addess}
%}

\maketitle

\begin{abstract}
\textit{Context:} Agile IT organizations, which are characterized by self-organization and collaborative social interactions, require motivating, efficient and flexible work environments to maximize value creation. Compressed work schedules such as the four-day workweek have evolved into multiple facets over the last decades and are associated with various benefits for organizations and their employees. \textit{Objective:} Our objective in this study is to deepen our comprehension of the impact of compressed work schedules on the operational efficacy of IT enterprises, while concurrently developing a comprehensive framework delineating the intricacies of compressed work schedules.\textit{Method:} We conducted a systematic review of available conceptualizations related to four-day workweek schedules and elaborate on their organizational and social effects. To cover scientific and practice-oriented literature, our review combined a systematic literature review and a web content analysis. \textit{Results:} Based on the generated insights, we derive a meta-framework that matches conceptualizations and effects, finally guiding the adoption of compressed work schedules based on individual managerial prerequisites and circumstances. 
\end{abstract}


\section{Introduction}
\label{Sec1:Intro}
In his 1930 essay ``Economic possibilities for our grandchildren," economist John Maynard Keynes predicted that his grandchildren's generation would work only three hours a day or 15 hours a week~\cite{Keynes.2010}. As far-fetched as this bold prediction may seem in retrospect, it reflected the trend at the time toward shorter working hours. In contemporary organizational landscapes, characterized by the integration of modern management methodologies and the disruptive influence of events like the COVID-19 pandemic, alongside shifts in workforce demographics, employees increasingly seek work environments conducive to operational flexibility, personal development, genuine interpersonal connections, and the harmonization of work and personal life. At the same time, particularly agile IT organizations are urged to shift their attention from the mere focus on functional targets to a perspective that emphasizes the exploitation of their employees’ potential in order to capitalize on the new nature of work and establish sustainable competitive advantage. The widespread adoption of technological innovations within managerial practice entails both, new risks and challenges as well as ingenious opportunities for organizations and their employees. Thus, modern companies can benefit from dispensing old industrial schedules, offering opportunities that enable their workforce to accomplish a sustainable work-life balance, and empowering employees to work smarter instead of longer. 

Accordingly, shortened work schedules have been established by many companies in recent decades (e.g., \cite{Kulak.2020,Pierce.1992}) to accommodate societal changes such as the growing number of dual-career households and sustain higher productivity and better quality. Four-day workweeks can increase the well-being, satisfaction, motivation and commitment of employees and are linked with several functional benefits such as an increase in efficiency and a decrease in fluctuation~\cite{Chakraborty.2007}. Yet, they also pose distinct challenges in terms of social collaboration, communication, and work organization. Such a consideration plays a particularly important role for agile IT companies, as agile working principles are characterized by a high degree of social interaction and temporal obligations. 

\begin{table*}
 \caption{Related work overview}
  \label{tab1:OverviewofRelWork}
  %\begin{tabular}{|p{0.15\linewidth}|p{0.25\linewidth}|p{0.55\linewidth}|}
   \begin{tabular}{ccp{0.58\linewidth}}
\hline
Publication & Research design & Brief summary \\
\hline
Campbell (2023)~\cite{Campbell.2023} & Chronicle systematic review & Based on a chronological systematic review, we examine how the evident effects of the 4-day week have changed over time in the literature. The author concludes that certain benefits persist and can be proven on a recurring basis. Examples include improved job satisfaction and motivation. On the other hand, negative effects are described, especially with regard to time-related aspects. With regard to performance, the review presents partly contradictory findings in the literature. \\
\hline
Munyon et al. (2023)~\cite{Munyon.2023} & Case study & The authors present best practices for a successfull application of a four-day work week in a case company. The aim of the study is to proof the evidence presented in literature based on their own empirical findings.\\
\hline
Veal (2023)~\cite{Veal.2023} & Systematic review & Focussing on leisure studies, the author analyzes the effects of a four-day work week on the employees leisure quality. Based on the results from a systematic review from different timeframes (in terms of decades), he analyzed the literature findings to identify how proponents of the four-day work week concept react to criticism. Finally, he concludes that considering the importance of social relevance and leisure quality, the effects of the four-day work week should be critically evaluated.\\
\hline
\end{tabular}
\end{table*}

For example, consider agile methods using retrospectives and review meetings. If a five-day workload is to be handled within four days, these practices are affected by the four-day workweek in a sense that they need to be conducted in a more time-efficient way (e.g., by streamlining the logging process) in order to minimize their time requirements and consequently maximize the time of the team members for rather output-related tasks (i.e., productivity)~\cite{Topp.2022}. Concurrently, team meetings and communication tasks need to be planned ahead of time to indemnify the availability of relevant team members, which is particularly important when employees can flexibly select a day off within a four-day workweek concept. Nevertheless, the adoption of reduced weekly working hours can sustainably strengthen the advised dual performance culture and enable agile IT organization to force an ambidextrous culture and face the dynamics of evolutionary and disruptive changes in the corporate environment in a context-appropriate manner~\cite{Lindskog.2021}.

With the rising prevalence of alternative work schedules within IT organizations, there is a pressing need for research to furnish practical, evidence-based insights into the circumstances under which shortened work schedules prove advantageous. While the potential implications for the work force have been extensively researched~\cite{Brough.2010}, there is a lack of corresponding insights regarding the individual conceptualizations and their particular effects on constituent organizations and employees. Accordingly, based on a systematic review of four-day workweek conceptualizations, this paper develops a meta-framework that delineates shared attributes and disparities among different conceptualizations and outlines their implications on the firm and employee level.

This paper is structured as follows: In Section~\ref{Sec2RelWork}, we provide an overview of related work. In Section~\ref{Sec3:Research Design}, we elaborate on our research design and introduce our research questions. Next, we present  the results of our study in sections~\ref{Sec4:Results_Concepts} and~\ref{Sec5:Results_Meta-Framework}. Finally, we outline the limitations of our study in Section~\ref{Sec6:Limitations} and conclude with a short discussion in Section~\ref{Sec7:Conclusion}.

\section{Related Work}
\label{Sec2RelWork}
Despite the increasing relevance of compressed workweek schedules and the growing scholarly interest in this topic, research in this domain is still scarce. Apart from studies that adopt a holistic focus on general advantages, disadvantages, opportunities, and challenges related to shortened workweeks (e.g., ~\cite{Brough.2010}), little research has been conducted to examine their conceptualizations and effects. Based on our research objectives and design (see Section~\ref{Sec3:Research Design}), Table~\ref{tab1:OverviewofRelWork} provides an overview of releated secondary and tertiary studies that deal with implications and success factors of compressed work schedules in managerial practice. 

%Based on our research design (see Section~\ref{Sec3:Research Design}), we focused primarly on secondary and tertiary studies. We conducted the search using Google Scholar, as we did not want to systematically exclude gray literature due to the small number of related works. Our search considered the last decade by applied year range filter. First and foremost, we found that literature in the area of compressed work schedules and the four-day work week in particular is growing. However, we were able to identify only a few papers related to our study. Table~\ref{tab1:OverviewofRelWork} gives an overview of the identified studies. 

The study that seems closest to ours in terms of methodology and objective is Campbell's chronicle systematic review~\cite{Campbell.2023}. However,  motivation and objectives are significantly different compared to our research. Campbell's study covers a long period of time, but does not aim to systematically identify different approaches of the four-day week in order to describe its high variety. Moreover, it adopts a generic focus rather than focusing on the particular case of (agile) IT organizations. This nuanciation is particulalry important, since  agile IT organizations such as software development companies emphasize communication and collaboration to react to highly dynamic business environments and create customer value. Since agile companies typically engage in environments that are characterized by high degrees of change and complexity, managers have to deal with multiple uncertainties. However, demanding certainty in the face of uncertainty is dysfunctional, which means that hierarchical command-and-control leadership styles are not suitable to agile work environments. In line with these distinct characteristics and immanent requirements, agile IT organizations feature a unique business environment that can be highly nourishing for the implementation of compressed work schedules such as the four-day workweek, while also featuring distinct challenges that need to be accounted for.

Nevertheless, Campbell's discussion offers a broader perspective compared to Veal's~\cite{Veal.2023} restrictive definition, which confines the 32-hour week model strictly to a four-day workweek. Comparable observations apply to the research design by Munyon et al.~\cite{Munyon.2023}, who, through a single case study, seek to identify best practices for the successful implementation and application of a compressed workweek. 

Overall, much of the foundational knowledge underpinning discussions and assumptions in studies within this area originates from a period preceding digitalization\cite{Campbell.2023} and its profound impact on work practices (e.g., working from home/remote work~\cite{Neumann.2023}), as well as the increased networking and market volatility~\cite{Bennett.2014} that have driven the widespread adoption of agile methods~\cite{Williams.2010}. 

To the best of our knowledge, our study is the first to analyze 4-day workweek concepts based on a systematic review and to evaluate the impact of these models on organizational performance in agile IT organizations. 

\section{Research Design}
\label{Sec3:Research Design}
 To extract different conceptualizations of the four-day workweek and grasp universal as well as concept-specific effects, we conducted a systematic literature review (SLR) as well as a web-content-analysis (WCA). The SLR has been conducted in line with the guidelines of Kitchenham and Charters~\cite{Kitchenham.2007}. The research process opted to provide a representative set of papers from the body of related publications to summarize the existing evidence concerning four-day workweek concepts and their organizational effects. We provide an overview of our search strategy and the study selection criteria in Table~\ref{tab:SearchStrategy}.
 
 In total the SLR process comprised three phases: (1) a planning phase that entailed the specification of guiding research questions (i.e., What conceptualizations of the four-day workweek are being used in organizational practice? What effects have been identified in the literature as a result of the four-day workweek? What effects are specific to individual conceptualizations of the four-day workweek?) and the development of a review protocol, (2) an execution phase that encompassed the selection, analysis and synthesis of relevant sources, (3) as well as a reporting phase. The keyword search (phase 2) was conducted between August and November 2023 for the fields abstract, title, and keywords using the academic search database Scopus. We excluded 1 search result that did not cope with our structural inclusion and exclusion criteria (i.e., written in English) as well as 42 results that did not fulfill the content-related criteria (i.e., managerial context of the investigation), finally resulting in a set of 17 papers. To examine content-related criteria, a sequential approach was adopted. Here, title, abstract, and keyword were examined first, before we proceeded to screen the introduction as well as conclusion and finally the full text of each article. Based on this procedure, the individual relevance (i.e., coding) of the search results was determined by assessing whether a publication is concerned with one or more conceptualizations of the four-day workweek in a managerial IT context. In this step, we excluded papers that reference four-day workweeks in non-managerial contexts (e.g., healthcare) or that refer to compressed  schedules that fall out of the scope of four-day workweeks (e.g., three-day workweeks). Ultimately, we identified 17 publications that were used to synthesize four-day workweek conceptualizations and their effects. The coding was validated by calculating the interrater reliability, yielding an agreement percentage of 100\% for structural and 97.88\% for content-related coding attributes. Furthermore, also the interrater agreement between the authors is very high, featuring agreement levels calculated by means of Cohen’s Kappa and Krippendorf’s Alpha between 0.9245 and 0.9246. In case of disagreements, the authors discussed their opinions to reach a joint verdict about the individual coding.

 \begin{table}[h!]
\caption{Search and selection strategy}
\begin{center}
\begin{tabular}{ | m{2.97cm} | m{5cm}| } 
  \hline
  \textbf{Database}& \textbf{Scopus}  \\ 
  \hline
  \textbf{Scope} & Primary/secondary studies and gray literature (e.g., company reports, academic writings and preprints)\\
  \hline
  \textbf{Search Term} & (4-day work* OR four-day work* OR compressed work*) AND (business OR organization OR organisation OR company)  \\ 
 \hline
 \textbf{Search fields} & Abstract, title, and keywords \\
  \hline
  \textbf{Inclusion criteria} & Studies that are … \\
  \hline
  Structural         & … written in English \\
                     & … dealing with four-day workweeks in organizational practice \\
  Content-related    & … outlining two or more characteristics of one or more four-day workweek conceptualizations \\
  \hline
  \textbf{Exclusion criteria} & Studies that are … \\
  \hline
  Structural         & … not written in English \\
  Content-related    & … related to schedules that do not concern four-day workweeks \\
                     & … published in non-managerial contexts \\	
\hline
\end{tabular}
\label{tab:SearchStrategy}
\end{center}
\end{table}
 
 Following Kim and Kuljis~\cite{Kuljis.2010}, we paralleled the methodology of the WCA to that of the SLR. To gather a sample of relevant conceptualizations, we used 'Google News' as search database and screened journalistic articles that were published since 2017. Search term as well as inclusion and exclusion criteria were set in accordance with the SLR process. In the WCA stage, we synthesized the insights from a set of 39 publications. Again, we validated our coding based on structural as well as content-related criteria, yielding an interrater reliability of 100\% (structural) and 99,2\% (content-related) as well as an interrater agreement between 0.9462 and 0.9463 for Cohen’s Kappa and Krippendorf’s Alpha.
 
 The result sheet of the review process including the all references of the SLR and WCA are available at Zenodo~\cite{DataExtract.2022}.

\section{Conceptualizations and effects of the four-day workweek}
\label{Sec4:Results_Concepts}

\begin{figure}[ht!]
    \caption{Characteristics of four-day workweek concepts}
    \centering
    \includegraphics[width=0.45\textwidth]
    {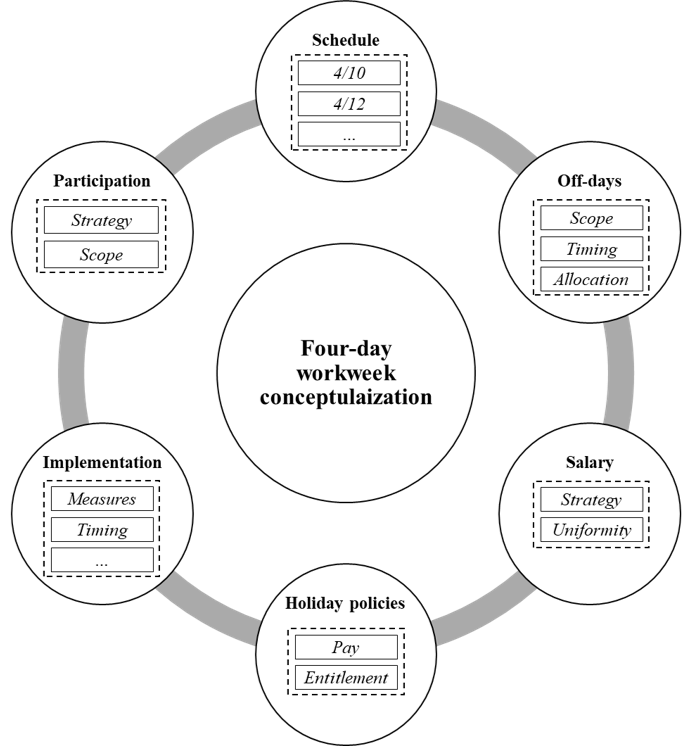}
    \label{fig:characteristics_4dwwconcepts}
\end{figure}

In practice, there is a multiplicity of different schemes that relate to four-day workweek schedules, each featuring distinct characteristics (e.g., ten hours per day, flexible off days, unmodified salary and mandatory participation). We refer to these schemes and their individual characteristics as conceptualizations and, in the following, outline their properties to structure the field of four-day workweek design. In principle, a conceptualization contains six characteristics, namely (1) schedule, (2) off-days, (3) salary, (4) holiday policies, (5) implementation, and (6) participation (see Fig.
\ref{fig:characteristics_4dwwconcepts}). 

\textbf{Schedule: }Schedule is the most apparent identifier of compressed work schedules and is concerned with the ratio between weekly working days and daily/weekly working hours. During our review, we found 11 schedules that are operationalized in practice and fall under the scope of a four-day workweek. Table~\ref{tab:fordayworkweekschedules} synopsizes these schedules.

\textbf{Off-days:} Off-days relate to the scheduling of non-working time that results from eliminating the fifth working day. Opportunities to approach the selection of off-days come down to three major dimensions, namely the scope (i.e., half-days or full days) and the timing of the off-days (i.e., fixed day, fixed rotation, or flexible/periodical selection), as well as the allocation of the off-day selection (i.e., organization, employee). From the interplay of these dimensions, several opportunities for designing off-day schedules are available for decision-makers, such as determining Fridays as fixed off-days (i.e., scope: full day; timing: fixed day; allocation: organization) or allowing the workforce to select two half off-days on a weekly basis (i.e.,  scope: half day; timing: flexible selection; allocation: employee).

\textbf{Salary:} Salary describes the monetary compensation that is provided to the workforce after implementation of the four-day workweek. Basically, companies can choose to follow two strategies: Maintaining the salary level of the five-day working week or adapting (i.e., reducing or increasing) salary payments based on the actual hours worked. Moreover, in terms of uniformity, firms can choose to apply the selected strategy uniformly across the entire organization, or adapt is for selected departments or business units.  

\textbf{Holiday policies:} Holiday policies relate to changes in holiday pay and entitlement regulations. Again, organizations can opt to maintain prior standards or adapt policies in accordance with new work design. Concerning holiday pay, paid time off can be calculated on an hourly basis instead of a daily, weekly, or monthly basis. In terms of entitlement, companies may provide employees with options such as accumulating vacation time or choosing an alternative day off in weeks where public holidays coincide with their regular off-day. 

\textbf{Implementation: }Implementation delineates the procedure that is adopted to put four-day workweeks into operational practice. Essentially, the implementation process can be characterized by the presence of implementation measures (e.g., employee surveys or interviews to estimate the fit to the company’s corporate culture; performance audits after adoption), post-implementation performance assessment timing (e.g., continuously or upon completion of certain milestones), type of implementation (i.e., time restricted pilot project or permanent shift), implementation timing (i.e., periodical/seasonal or year-round), and transition (i.e., adopting a transitions phase such as 4.5 working days a week or directly shifting towards the envisaged four-day workweek schedule). 

\textbf{Participation:} Participation is concerned with the actions and incentives taken to involve employees in the new workweek design. Here, companies can choose a strategy that makes participation mandatory or leave the choice up to their employees (i.e, voluntary). In this context,  mandatory and voluntary participation principles can differ in scope, meaning that they either relate to the entire workforce, or selected teams/employees whose workflows seem particularly suitable for the adoption of shortened workweeks.

Based on the combination of different parameters, four-day workweek conceptualizations can have remarkably different implications on organizations and their employees (see Section~\ref{Sec5:Results_Meta-Framework}). In a nutshell, they can be broken down into organizational and employee-related effects. Organizational effects directly relate to performance metrics of a company, whereas employee-related effects refer to well-being and attitude of the workforce. Moreover, these effects can be clustered based on the type of impact, namely whether they exert a positive or negative influence on a  firm or its employees. Table~\ref{tab:effects_4dwwschedules} presents impacts that have been extracted during our review process. While some effects seem to be universally positive or negative across all concepts, it is important to note that, depending on the individual conceptualization, an effect can also have both, a positive as well as a negative effect on organizations or its employees (as marked with an * in Table~\ref{tab:effects_4dwwschedules}). 

\begin{table}[h]
\caption{Four-day workweek schedules}
\centering
\begin{tabularx}{8.9cm}{|c|c|X|}
\hline
\textbf{Schedule} & \textbf{Type} & \textbf{Description} \\
\hline
3-4/12 & Alternating & Periodic (e.g., weekly) shift from three to four workdays with 12 hours per day (fixed) \\
\hline
4/10 & High-load & Four workdays with 10 hours per day (fixed) \\
\hline
4/12 & High-load & Four workdays with 12 hours per day (fixed) \\
\hline
4/24 & Low-load & Four workdays with 24 hours per week (daily working hours vary from 5.5 to 6.5 hours) \\
\hline
4/30 & Low-load & Four workdays with 30 hours per week (daily working hours vary from 7 to 8 hours) \\
\hline
4/32 & Low-load & Four workdays with 32 hours per week (daily working hours vary from 7.5 to 8.5 hours) \\
\hline
4/35 & High-load & Four workdays with 35 hours per week (daily working hours vary from 8 to 9.5 hours) \\
\hline
4/36 & High-load & Four workdays with 36 hours per week (daily working hours vary from 8 to 10 hours) \\
\hline
4/38 & High-load & Four workdays with 38 hours per week (daily working hours vary from 9 to 11 hours) \\
\hline
4-5/9 & Alternating & Periodic (e.g., weekly) shift from four to five workdays with 9 hours per day (fixed) \\
\hline
9/80 & Alternating & Weekly shift from four to five workdays (daily working hours vary from 8 to 10 hours). \\
\hline
\end{tabularx}
\label{tab:fordayworkweekschedules}
\end{table}

On the organizational level, positive as well as negative implications of four-day workweek schedules can be related to costs, employment, performance, and work environment. 

\textbf{Costs:} Positive implications in this cluster include the reduction of power and other utility costs (e.g., water), as well as human resource (HR) costs, including overtime payments, which result from a decreased utilization of infrastructure and reduced working hours. Yet, for some conceptualizations that are accompanied by a relative increase of holidays or sick days, which results from a decrease in working days, or additional overtime payments that are legally required in some countries, HR costs can also increase. Similarly, some organizations require high availability teams (e.g., customer service), which may need to compensated for their extended availability (compared to the schedules of other departments), which also leads to an increase in HR costs in this category.

\begin{table}[h]
\caption{Effects of four-day workweek schedules}

\centering
\begin{tabularx}{\linewidth}{|c|c|X|X|}
\hline
\textbf{Type} & \textbf{Direction} & \textbf{Organization} & \textbf{Employees} \\
\hline
\multirow{2}{*}{Positive (+)} & Decrease $\downarrow$ & Utility costs, Costs for childcare*, HR costs*, Employee fluctuation*, Absenteeism & Commuting costs, Stress*, Burn-outs, Commuting time \\
\cline{2-4}
 & Increase $\uparrow$ & Corporate attractiveness, Productivity, Service quality, Coordination effectiveness*, Quality of infrastructure & Health and well-being, Moral*, Motivation, Engagement, Job satisfaction*, Life satisfaction, Leisure time*, Work-life balance*, Job autonomy, Work performance*, Work efficiency* \\
\hline
\multirow{2}{*}{Negative (-)} & Decrease $\downarrow$ & Task coverage & Moral*, Job satisfaction*, Work-life balance*, Leisure time*, Social interactions, Work efficiency*, Work performance* \\
\cline{2-4}
 & Increase $\uparrow$ & HR costs*, Employee fluctuation*, Coordination effectiveness*, Accounting efforts & Costs for childcare*, Physical fatigue, Mental fatigue, Stress* \\
\hline
\multicolumn{4}{|p{\dimexpr\linewidth-2\tabcolsep}|}{* Effect can feature a positive and negative impact based on context} \\
\hline
\end{tabularx}
\label{tab:effects_4dwwschedules}
\end{table}

\textbf{Employment:} Implementing four-day workweeks is likely to increase the corporate attractiveness for potential employees, which supports recruiting processes for both, high- and low-potentials. Similarly, it can have a remarkable influence on the workforces’ job satisfaction and thus decrease employee fluctuation and absenteeism. Still, some concepts that entail a reduction in salary and corporate benefits such as holiday pay, employee fluctuation may also be increased. Additionally, four-day workweek schedules create a work environment that holds immanent barriers to part-time jobs, as four-day workweeks restrict the flexibility for scheduling part-time.

\textbf{Performance:} In terms of performance, four-day workweeks have been proven to be capable of decreasing the rate of missed deadlines, which can mainly be attributed to the fact that these schedules foster a culture of thorough operational and strategic planning. Furthermore, resulting from an improvement of the priory mentioned planning culture as well as an increase in job satisfaction and autonomy, many organizations have seen an amendment in productivity, service quality and coordination effectiveness. On the contrary, some four-day workweek schedules can lead to a decrease in task coverage, meaning that the overall number of job tasks that can be handled by the workforce declines due to the reduction of working hours. Moreover, four-day workweeks with flexible timing can   entail a reduction in coordination effectiveness, as they raise availability issues among teams and departments. 

\textbf{Work environment:} Finally, four-day workweeks can benefit the work environment by improving the quality of the physical infrastructure because organizations have more temporal flexibility for maintaining and repairing assets such as computers. On the negative side, schedules that adapt salary and holiday payments based on temporal measures are concomitant with soaring efforts for payroll accounting. 

On the employee level, effects can refer to costs, health, attitude, private life, work environment and performance. 

\textbf{Costs:} Positive effects for employees include decreasing costs for childcare, which follow from an increasing flexibility in scheduling private activities, as well as reducing commuting costs, as employees need to visit an organization’s office or production facility one day less per week. Nevertheless, shortened workweek schedules that result in multiple over-hours due to longer workdays may also negatively influence childcare costs, as employees need to appoint an external assistant to take care of their children for these over-hour days.

\textbf{Health:} Concerning health-related benefits, four-day workweeks have proven to decrease the perceived stress level and reduce the occurrence of burn-outs, while sustainably increasing the overall well-being of the workforce. However, in this context it is important to note that many companies and their employees report an increased stress level as well as a decreased level of well-being during the first month after the implementation of compressed work schedules, which is then replaced by a significant decrease in perceived stress as well an increase in well-being after the workforce has got used to the new work design. Still, managers and companies should not underestimate that especially concepts, which involve an overall equal or additional workload can also increase the mental and physical fatigue level of their employees. 

\textbf{Attitude:} When it comes to the (work) attitude of employees, four-day workweeks hold the potential to increase moral (i.e., in a sense of positive attitude towards job and employer), motivation (i.e., in a sense of intrinsic intention to work efficiently as well as beneficial to the company), engagement (i.e., in a sense of being willing to engage beyond the normal scope of work), and job satisfaction (i.e., in a sense of feeling enjoyment about work duties and environment). However, moral and job satisfaction in particular seem to be contingent on the implementation scope. Accordingly, in cases where four-day workweek schedules are established only partially in a given company, these two variables are likely to experience a significant decrease for employee groups that are not granted to adopt a compressed workweek schedule.

\textbf{Private life:} Concerning the private living conditions of employees, four-day workweeks are likely to reduce commuting times, increase the staff’s amount of leisure time, and improve the work-life balance and overall life satisfaction. Yet, again, leisure time and work-life balance still depend on the particular concept of the four-day workweek, since schedules with high amounts of daily working hours may also be perceived as negative for the given leisure time and work-life balance (i.e., because they reduce the daily time for leisure activities, family, and children when employees return home late).

\textbf{Work environment:} The evidence of various companies has shown that four-day workweeks positively impact the job autonomy, meaning that employees receive more freedom to execute their job duties in a self-organized manner. The reason for that lies in the fact that shortened workweeks foster an environment of velocity and efficiency, which is necessary to complete given duties within four work-days. On the negative side, however, this type of environment naturally ensures that social interactions are minimized, as these would be counterproductive to the efficient and fast-paced environment.

\textbf{Performance: }Ultimately, compressed work schedules also affect the output of the workforce, namely the performance (i.e., velocity, service/product quality, error-rate) and efficiency (i.e., ratio of outputs to time that is required to achieve these outputs). As for other variables, both of these metrics can also be negatively influenced by four-day workweeks, when the respective schedules create an increased level of mental and/or physical fatigue in the affected employee groups. 

%\begin{framed}
%\noindent\textbf{Key Take Aways: } TODO! 
%\end{framed}

\section{A meta-framework for the adoption of four-day workweek schedules}
\label{Sec5:Results_Meta-Framework}
Having discussed the different conceptualizations and effects of four-day workweek schedules, the question remains, how the individual design of a schedule affects its implications, or, in other words, how to match the four-day workweek concepts with the set of organizational and social consequences.

\begin{figure}[ht]
    \centering
    \includegraphics[width=0.46\textwidth]
    {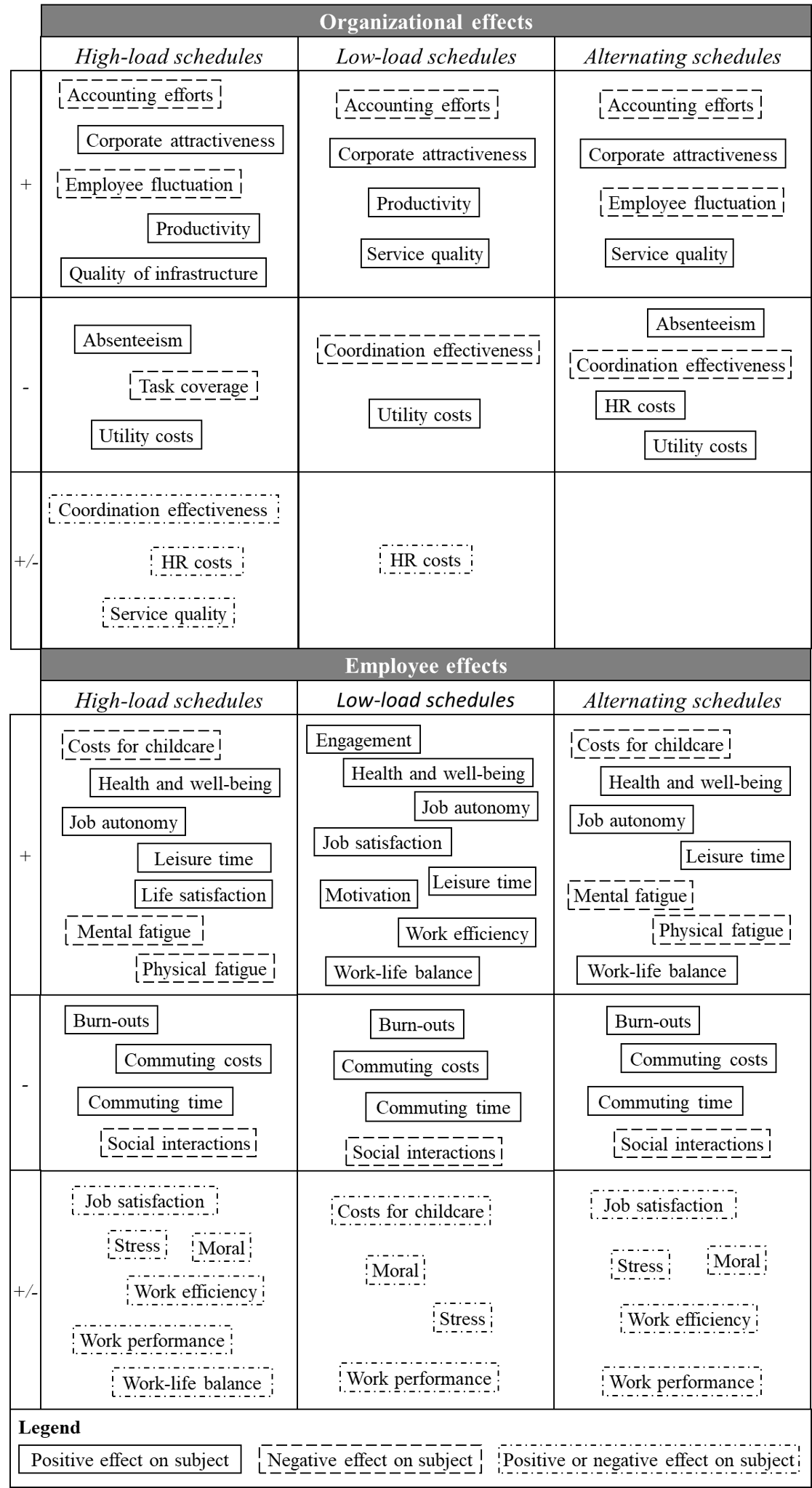}
    \caption{Meta-framework of four-day workweek effects}
    \label{fig:meta-framework-4dwweffects}
\end{figure}

\begin{figure}[ht]
    \centering
    \includegraphics[width=0.4\textwidth]
    {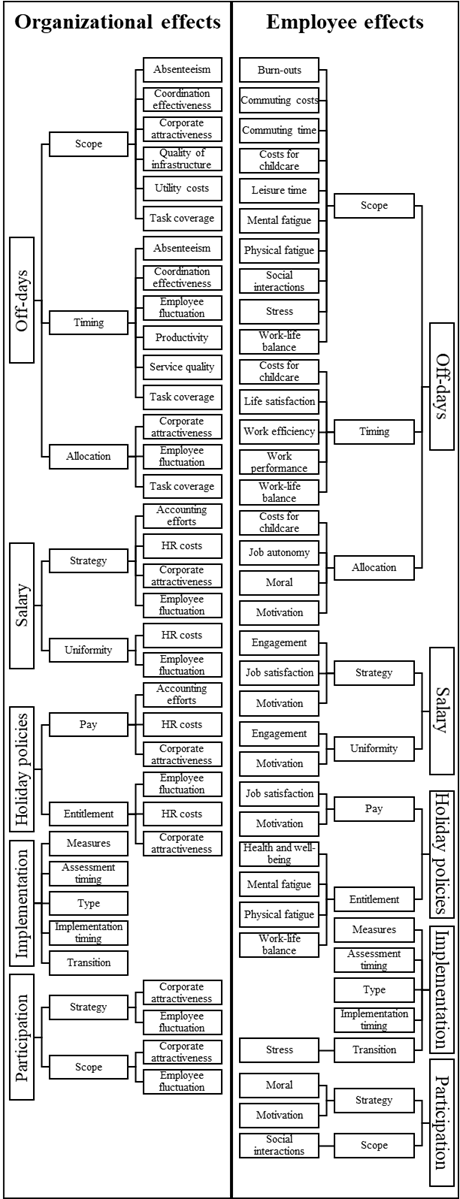}
    \caption{Relationship between four-day workweek concepts and effects}
    \label{fig:framework-4dwweffects}
\end{figure}

As outlined in Fig.~\ref{fig:meta-framework-4dwweffects}, high-load (e.g., 4/10), low-load (e.g., 4/30) and alternating (e.g., 9/80) work schedules have different effects on organizational and employee-related metrics. Concerning organizational implications, all schedules entail an increase in accounting efforts and corporate attractiveness, as well as a decrease in utility costs.  However, high-load and alternating schedules seem to be less beneficial in terms of employee fluctuation. While high-load and low-load schedules positively affect productivity, this is not the case for alternating schedules, as these do not allow a consistent switch towards compressed schedules, thus impairing a sustainable development of more efficient work patterns. Service quality can benefit from all schedules, but is contingent on other factors such as timing of the off-days in the case of high-load factors (see also Fig.~\ref{fig:framework-4dwweffects}). This can mainly be explained by the fact that high-load schedules entail high workloads, which in turn result in an increase in mental and physical fatigue of the workforce. In-low load and alternating schedules, the overall stress to which employees are exposed is lower, so that they can perform better and increase they quality of the services they offer. On the contrary, depending on the off-day timing, coordination effectiveness can even increase for high-load schedules, while it is likely to decrease in low-load (e.g., due to limited availability of employees) and alternating schedules (e.g., due to the lack of consistent planning).

Concerning employee-related effects, all schedules are in favor of the workforce’s health and well-being, the occurrence of burn-outs, job autonomy, (perceived) leisure time, as well as commuting costs and efforts (i.e., in terms of time spend). Yet, they also account for negative effects, such as a decrease in social interactions at work, which might for example impair moral and job satisfaction. Not surprisingly, high-load and alternating schedules also result in an increase in (perceived) mental and physical fatigue, which are directly related to an employee’s work efficiency (in terms of input to output ratio) and performance (in terms of output quality). In line with the decreased workload, low-load work schedules particularly improve motivation, engagement and work-life balance, whereas their effects on childcare costs mainly depend on off-day scope, timing, and allocation. When employees are allowed to flexibly plan off-days in accordance with their personal needs, childcare costs are likely to decrease since they can organize themselves in a way that avoids the need for external childcare services. In contrast, high-load and alternating schedules generally seem to increase childcare costs, because the (partial) need to work over-hours restricts the workforce in its ability to organize childcare privately. Correspondingly, work-life balance is also not directly improved by high-load schedules, but depends on individual scope and time of the off-day planning as well as the given regulations regarding holiday entitlement. Other factors that are not  influenced directly by the work schedule are job satisfaction and work efficiency in the case of high-load and alternating schedules as well as stress, moral, and work performance in the case of all schedules.  
To additionally support managers in adapting conceptual factors that might support or counter the effects of a given schedule, Fig.~\ref{fig:framework-4dwweffects} presents an overview of the relationship between conceptual identifiers and the potential effects of four-day workweeks. On the organizational side, the scope, timing, and allocation of off-day planning can have an remarkable impact on various metrics. For example, absenteeism reduces with full off-days (i.e., scope) that can be planned flexibly (i.e., timing) by the employees, as it allows to combine private needs with work. However, half-days (i.e., scope) and fixed planning (i.e., timing) minimize these advantages and will not have a major impact on absenteeism. Similarly, employee fluctuation can benefit from flexible off-day planning (i.e., timing), staff-based off-day selection (i.e., allocation), unitary (i.e., uniformity) maintained salaries (i.e., strategy) as well as holiday regulations (i.e., entitlement), and voluntary (i.e., strategy) participation in the new schedules across the entire company (i.e. scope). Nevertheless, fluctuation might also increase with the implementation of four-day schedules when these entail a decrease in salary (i.e., strategy) and a reduction in holiday entitlement or when they are only implemented for selected departments (e.g., scope of participation).

On the employee level, relevant interdependencies include the relationship between off-day scope and fatigue (i.e., as full days reduce fatigue  stronger than half days), off-day allocation and job autonomy (i.e., as being allowed to schedule off-days in a self-contained manner provides a greater sense of self-governance), and participation scope and social interactions at work (i.e., as companies that uniformly adopt four-day work schedules are more likely to establish an overarching sense of task-dedication and work efficiency). Additionally, transition phases can have a remarkable effect on an employee’s perceived stress level during and after the implementation of four-day workweeks, because they help the workforce to culturally and mentally adapt to the new work design. Indeed, our review has shown that stress is primarily a relevant factor for workers within the initial period after the introduction of a four-day workweek, and that it seems to decrease steadily as employees become more accustomed to this concept.

Finally, it is important to mention that the presented effects are not solely contingent on the respective schedule, but are the result of the individual interplay of work design, workforce characteristics, and organizational context. Nevertheless, the proposed framework offers a valuable starting point to identify implication-related differences and commonalities of four-day work schedules and enables IT managers to grasp the directs effects of certain schedules (Fig.~\ref{fig:meta-framework-4dwweffects}) as well as the inverse contingencies with other metrics (Fig.~\ref{fig:framework-4dwweffects}).

\section{Limitations}
\label{Sec6:Limitations}
As explained in Section~\ref{Sec3:Research Design}, we decided to design the systematic review by using a combined approach of a SLR and WCA. We conducted our study based on the rigorous guidelines of Kitchenham and Charters~\cite{Kitchenham.2007} and Kim and Kuljis~\cite{Kuljis.2010}. Still, like any research, our study features somt limitations that need to be considered.

One of the main challenges in conducting systematic reviews relates to the completeness of the identified result set applying the inclusion and exclusion criteria. To mitigate the potential omission of essential studies, we conducted test searches in Scopus and Google Scholar, which yielded highly redundant results. While several studies endorse Scopus as a suitable standalone database for secondary studies (e.g., \cite{Alsaqaf.2017,Stray.2020}, we are aware of the risk not identifying all the relevant studies in the field. Another limitation applies to the selection process of the identified literature. In Section~\ref{Sec3:Research Design}, we provide a detailed explanation of the selection process and the quality assurance actions we took to mitigate the bias of the individual researchers involved in this study.
Finally, acknowledging the prominence of our research domain, characterized by widespread interest among researchers and practitioners, we have opted to incorporate a WCA into our systematic review. This decision is driven by our aim to encompass current insights and perspectives from the practitioner community, ensuring the inclusion of up-to-date knowledge.

\section{Conclusion \& Future Work}
\label{Sec7:Conclusion}
In today’s competitive and dynamic business environment, workplace practices are significantly changing. On the one hand, organizations are looking for employees who can quickly adapt to change and lead with agility and self-determination. On the other hand, employees demand working conditions under which they can fully realize their individual potential and establish a satisfactory work-life balance. Compressed work schedules like the four-day workweek hold the potential to add value on the business and employee level by leveraging a work design that meets the immanent requirements of the new workforce generations and by cultivating a culture that promotes efficiency, strategic planning, and knowledge accumulation. Particularly in agile IT organizations, which are characterized by high degrees of collaboration and self-organization, managers need to make every hour effective, so that employees can be become active co-creators and drivers of change, value, and innovation. 
Since they are a relatively uncommon work schedule practice, to date, four-day workweeks are employed only by a few organizations around the globe. As a result, organizations lack reliable information on shortened workweek designs and their potential impacts on organizational performance and employee satisfaction, which constitutes a major impediment to the adoption of four-day workweeks in practice. Our research has identified a set of conceptualizations and implications that can guide the decision to adopt four-day workweeks. Due to their distinct business characteristics, these insights can be particularly viable for agile IT organizations and agile development teams. Unfortunately, there is no single truth when it comes to the applicability and effects of compressed work schedules. In the end, the successful implementation of four-day workweeks depends on numerous contextual (e.g., management commitment) and structural factors (e.g., industry), which we seek to survey in future research. Still, we believe that the provided information on concept-dependent effects as well as the proposed meta-framework offer a valuable starting point to educate managers about the potential benefits of four-day workweeks and provide them with a structured approach to adopt shortened workweeks for the benefit of the company and its employees. 

\bibliographystyle{IEEEtran}
\bibliography{references}

\end{document}